# Optoelectronic Trajectory Reconfiguration and Directed Self-Assembly of Self-Propelling Electrically-Powered Active Particles

Sankha Shuvra Das[1] and Gilad Yossifon[1]*

[1]School of Mechanical Engineering, Tel-Aviv University, Tel-Aviv 69978, Israel

## Abstract

Self-propelling active particles are an exciting and interdisciplinary emerging area of research with projected biomedical and environmental applications. Due to their autonomous motion, control over these active particles that are free to travel along individual trajectories, is challenging. This work used optically patterned electrodes on a photoconductive substrate using a digital micromirror device (DMD) to dynamically control the region of movement of self-propelling particles (i.e. metallo-dielectric Janus particles (JPs)). This extends previous studies where only a passive micromotor was optoelectronically manipulated with a translocating optical pattern that illuminated the particle. In contrast, the current system used the optically patterned electrode merely to define the region within which the JPs moved autonomously. Interestingly, the JPs avoided crossing the optical region's edge, which enables constraint of the area of motion and to dynamically shape the JP trajectory. Using the DMD system to simultaneously manipulate several JPs enabled to self-assemble the JPs into stable active structures (JPs ring) with precise control over the number of participating JPs and passive particles. Since the optoelectronic system is amenable to closed-loop operation using real-time image analysis, it enables exploitation of these active particles as active microrobots that can be operated in a programmable and parallelized manner.

## Introduction

Self-propelling active particles are the focus of an exciting and interdisciplinary emerging area of research, with projected applications in drug delivery[1], detoxification[2], environmental remediation[3], immunosensing[4], remote surgery[5], self-repairing systems[6], and others. Motion is achieved by designing particles that can asymmetrically draw and dissipate energy, creating local gradients of force for autonomous propulsion, even under uniform ambient conditions. Externally imposed macroscopic gradients of the force fields, essential for the movement of passive particles, can lead to particle motion by effects such as dielectrophoresis[7] and magnetophoresis[8,9]. However, these all result in phoretic motion; i.e., migration en-masse in an externally dictated direction along the applied field gradient. In clear contrast, self-propelling particles are free to travel along individual trajectories[10], i.e., autonomous motion, which allows them to: (i) operate under simpler-to-realize field conditions (i.e., uniform ambient field) than phoretically driven particles; (ii) efficiently cover areas and perform tasks within closed microscale spaces; (iii) exhibit collective behaviour; and (iv) self-assemble into more complex active structures. Uniquely, under the application of a uniform AC electric field, we and others have shown self-propulsion of metallodielectric Janus particles (JPs) in either induced-charge electrophoresis (ICEP) or self-dielectrophoresis (sDEP) modes[10]. Importantly, we and others showed that varying the frequency of the applied electric field can give rise to a number of distinct electrokinetic effects[23] that can power locomotion in different ways[24], control the interaction of the active particle with other active/passive[25,26] particles (e.g., cells), and control the collective behaviour of many JPs[27].

However, there are cases where accurate control over the trajectory of an active particle is needed, e.g., interaction with a single targeted cell[11] or cargo delivery to a target destination[12], requiring an external mean of steering (e.g., magnetic). Our group recently demonstrated the electrical interaction between a single active particle and a targeted mammalian cell for both its local electroporation for injection of drugs/genes[13]. In addition, we demonstrated the possibility of in-situ electrodeformation of the cell nucleus[14], which can be used as a mechanical biomarker and avoids the need for externally invasive manipulators. However, these features were demonstrated for a manually controlled single active metallo-dielectric JP (one hemisphere has metallic coating). It is impossible to navigate several individual active particles independently, as they all respond to the single applied field that either propels or stirs all particles at once. Hence, their application is limited to microrobots made of either a single active particle[11] or a collective (microswarm[15,16,17]), resulting in low throughput operation and fixed configuration. The inability to simultaneously control several JPs can result in most becoming inoperative, due to self-interaction with other JPs to form random clusters/chains.

A platform that enables dynamic transition between phoretic and self-propelling motion in a programmable and parallelized manner will be able to achieve independent control of many active particles simultaneously and even to direct their self-assembly. This essentially necessitates spatial discretization of the applied field, as exists in an array of microscale electromagnets[18,19] for the manipulation of magnetically labelled objects or multiplexed optical tweezers for the manipulation of objects via optical forces[20]. However, while these are limited to phoretic manipulation, the optoelectronic tweezer (OET)[21] system is the most useful for our purposes, as it can serve to unify the operation of the system using a singularly controlled electric field that can control both phoretic and electrically driven self-propulsive motion. The OET can achieve the former via direct tweezing of particles and the latter by optically defining the virtual boundaries of the region within which the electric field exists and active particles can propel. It enables a shift from phoretic to self-propulsive motion of electrically-driven active particles, by appropriate increase of the illuminated area within which the active particle is restricted to move. It relies on light-induced DEP to manipulate microscale objects[22] and is well suited for parallel manipulation schemes by generating and controlling multiple light beams via a digital micromirror device (DMD) to optically pattern the electrodes in a dynamic and programmable manner. Other advantages are its relative ease of operation and simple microfluidic chip with a uniform photoconductive coating that becomes conductive upon illumination. The OET system may also be capable of controlling, via phoretic manipulation, the number and type of constituents within the illuminated region. In turn, this can guide self-assembly into specific conformations (e.g., see in Nishiguchi et al. 2018[28] the huge dispersity in the conformations of the stochastically self-assembled colloidal structures). Moreover, the system may combine non-active particles to form complex hybrid passive/active structures; e.g., a chiral spinner[29] or collective interaction with a cell[30], in a deterministic manner unlike the random assembly obtained in Refs.29 and 30. Sufficiently large JPs can act as active carriers that can manipulate (load, transport and release) cargo of synthetic/biological particles in a unified, label-free, selective manner, as we observed in our previous study.[26] While the OET platform has only been successfully applied for manipulation of passive micromotors[22], we aimed to study its combination with reconfigurable self-propelling micromotors. We studied the self-propulsion behaviour of single active particles and the dynamic configuration of their trajectories via optical patterning (Fig.1b). In addition, we studied reconfigurable self-propelling micromotors that can be dynamically assembled in a bottom-up approach by combining different active (and even passive) particles into any desired combination (number, type) and motion mode (self-propulsion, phoretic, collective) (Fig.1b).

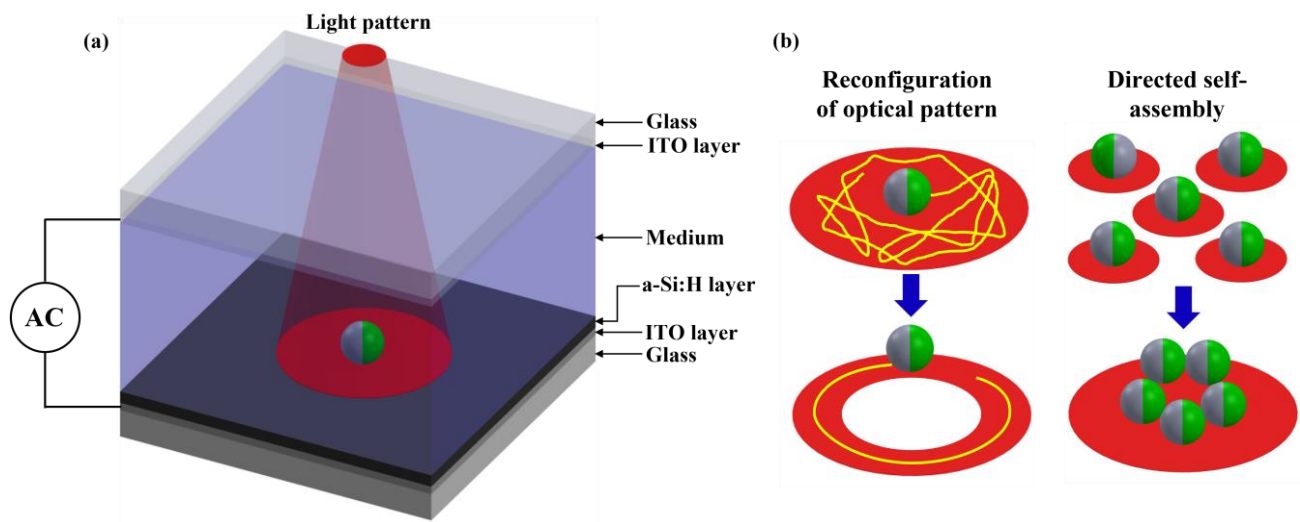

**Figure 1:** (a) Schematic of the microfluidic chamber for the optoelectronic (OE) manipulation of active particles; (b) Reconfiguration of active particle trajectories as well as direct self-assembly of active particles. Here, green and gray side of the active particle represents the metallic and dielectric part, respectively.

## Results and Discussion

### *Self-propulsion of a single JP within an optically patterned region*

Examination of the propulsion of a JP within an optically patterned circular region under varying frequencies found that under both low and high frequency regimes corresponding to ICEP and sDEP propulsion modes, respectively, the JP was moved within the optically patterned region only (Fig.2a). However, there were distinct differences between the low and high frequency motion of the JP within the optically patterned region. While in the high frequency regime the JP moved in circular path at some distance from the edge of the optical region, in the low frequency regime, the JP moved randomly within the entire region of the optically patterned region while keeping a larger distance from the edge of the optical region. Movement of the JP can be quantified in terms of the particle's mean square displacement (MSD) versus time behavior (see Fig. 2b), using the expression $MSD = \frac{1}{N-n}\sum_{i=1}^{N-n} l_{i,i+n}^2$, wherein $l_{i,i+n}$ is the displacement between two particle locations (i.e., $i$ and $i + n$), and $N$ is the total number of data points[31]. As the particle trajectory is measured over a fixed sampling time, $\Delta t$ (~O(35ms for Fig. 2b)), a specified time interval ($t$) over which MSD is evaluated, can be defined as the integer multiple of $\Delta t$ (or, $t = n\Delta t$). Therefore, for a particle trajectory with $N$ (=1681, for Fig. 2b) number of data points, $N - n$ describes the number of data pairs separated by the specified time interval (for e.g., 35 ms for n =1, 70 ms for n = 2, and so on). As expected, the ordered circular versus random trajectories of the ICEP and sDEP motion, respectively, exhibited a periodic versus random MSD time-dependent response (Fig. 2b).

Examination of the velocity versus frequency of the JP (Fig. 2c) identified the expected reversal (at ~60kHz) in the direction of motion of the JP[10,32], which transitioned from movement with its dielectric end forward (ICEP) to its metallic end forward (sDEP). Microscale particle image velocimetry (µPIV) was performed to better understand why the JP in both motion modes of ICEP and sDEP did not cross the edge of the optically illuminated region. The analysis clearly showed that a strong electrohydrodynamic flow (EHD) directed radially inward from the edge of the optical pattern was generated in the low frequency regime (10 kHz) (Fig.S2). Such EHD[33] or alternating-current-electro-osmotic (ACEO) flow[34] is a nonlinear electrokinetic phenomenon associated with the charge induced onto the powered electrodes and scales with the following RC (resistor-capacitor) time $\tau = \lambda H/D$. For a double layer length $\lambda \sim 42nm$ (50µM KCl), microchamber height $H$=120µm, ion diffusion coefficient $D = 2 \cdot 10^{-9} \ m^2/s$, the characteristic RC time corresponds to an applied frequency of approximately $f_{RC,ACEO} = 1/2\pi\tau \sim 64Hz$. Since the intensity of the ACEO velocity decays as $1/f^2$ [7] for $f \gg f_{RC,ACEO}$, it is expected to play a minor role in the high frequency regime (100kHz), in agreement with the reduction of flow velocity obtained in the µPIV analysis (Fig. S2) at 100kHz relative to 10kHz. This enables ordered JP motion in a circular path at 100kHz (Fig. 2a), suggesting the lack of fluctuations due to EHD (as in 10kHz). Such ACEO flow was also confirmed using numerical simulations (see Fig.S3 and Methods and Materials section), which showed a vortex pattern with an inward radial flow with a flow intensity that monotonically decays with increasing frequency beyond that of ~$5f_{RC,ACEO}$ (~320Hz). The distance of the JP from the edge of the optical pattern while under the sDEP mode may be explained by the fact that the optical intensity of our setup (Fig. S1) was non-uniform, especially near the edge (Fig. 2d), which suggests that the effective photoconductive electrode region is smaller than the apparent illuminated region. Since the JP exhibits positive dielectrophoretic (pDEP) behavior as was experimentally characterized using a quadrupolar electrode array[35] (Fig. S4), it is expected to be attracted to the line of maximum electric field intensity and gradient, which correspond to the location where the optical intensity reaches a sufficiently high value to assure that the illuminated region behaves as a conductor. Approximating this threshold value of optical intensity to induce sufficiently strong photoconductive effect as a normalized intensity value of 0.5-0.6 (Fig. 2d) suggests that the true edge of the effectively conducting region is 55-60μm away from the edge of the optical pattern (approximated to correspond to a normalized intensity value of 0.1), in qualitative agreement with the location of the sDEP path relative to the optical pattern's edge observed in Fig. 2a. A more quantitative measurement of the dependency of the a-Si:H photoconductive substrate's conductivity on the light intensity is depicted in Fig.S8, exhibiting a light intensity threshold for a measurable photoconductive effect.

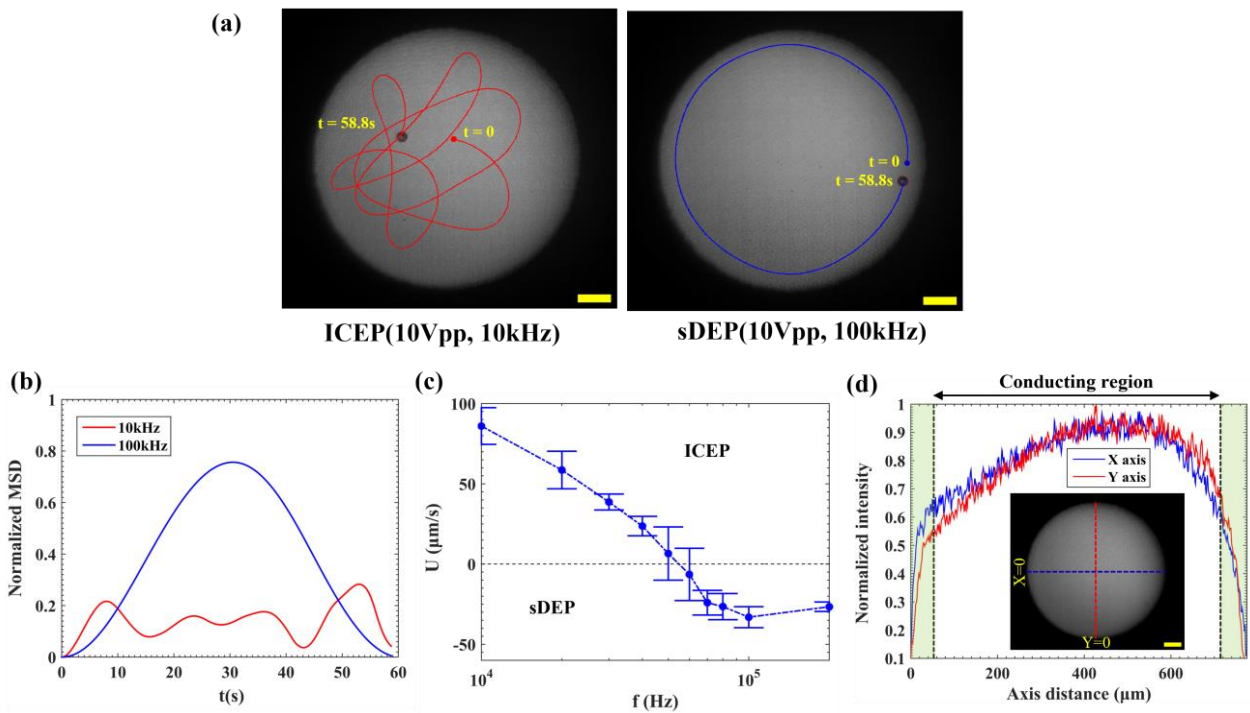

**Figure 2: Electrically powered self-propulsion of an active particle within an optically defined region onto a photoconductive substrate.** (a) Trajectories of a Janus particle (JP) (27 μm in diameter) in both ICEP (10 kHz) and sDEP (100 kHz) modes. See also supplementary video S1. (b) The temporal variation of the normalized MSD ($= MSD/D_B^2$, where $D_B$ ~770 μm is the beam diameter) of the JP self-propulsion behavior. (c) Frequency dispersion of JP propulsion velocity showing its direction reversal from ICEP to sDEP mode at 50-60 kHz. (d) Normalized intensity profile of the optically defined "beam" region along the x and y axes. Error bars represent standard deviation values of the propulsion velocity. A low conductive electrolyte of 50 μM KCl with 0.1% (v/v) Tween-20 (conductivity ~8 μS/cm, pH ~5.4) was used as the background medium. The magnitude of applied voltage was 10V$_{pp}$. Scale bars: 100 μm.

Further examination of the effect of beam size on JP propulsion of the JP found that the qualitative behavior of the ICEP and sDEP movement did not change when constraining the JP to move within a smaller beam size (Fig. 3a, b); in the former mode, the JP moved randomly within the beam while in the latter mode, it moved in a circular trajectory. However, increased constriction with decreased beam size significantly reduced the MSD. The mostly similar JP behavior despite different beam sizes, was also reflected in the measured JP velocities (Fig. 3c), where only the low frequency (10 kHz) velocity increased by ~20% for the small beam size. The latter may result from EHD-induced JP levitation,[32] which decreases the effect of the JP-wall interaction on the Stokes drag.

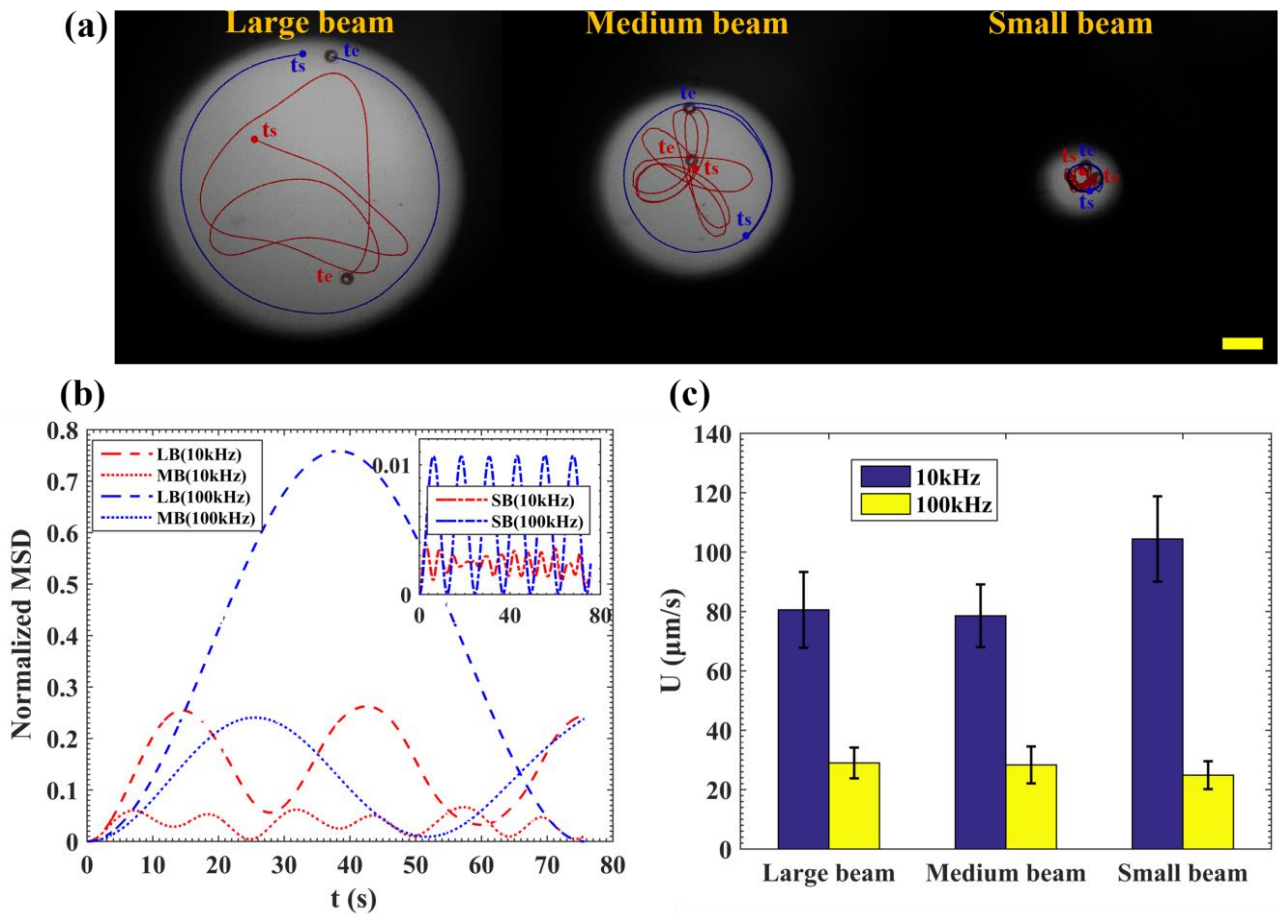

**Figure 3: The effect of the optically patterned region's size on the active particle self-propulsion.** (a) ICEP (10kHz) and sDEP (100kHz) self-propulsive behavior of a JP (27 μm in diameter) within beams of varying diameters (large beam (LB) 770 μm, medium beam (MB) 495 μm, small beam (SB) 180 μm). See also supplementary video S2. Scale bar: 100 μm. (b) The temporal variation of the normalized (by the illuminated beam diameter, i.e., $D_B^2$) MSD of the JP self-propulsive behavior for varying beam size. (c) Average propulsion velocities of JP in both ICEP and sDEP modes for different beam sizes. Error bars represent standard deviation values of the propulsion velocity. An applied voltage of 10V$_{pp}$ was used with a low-conductivity electrolyte of 50 μM KCl with 0.1% Tween-20.

*Translocation and directed self-assembly of JPs*

With the aim of directing self-assembly, we first examined the ability to translocate the JP by reconfiguring the optical pattern instead of moving the illuminated region and dragging the JP along with it. As clearly shown in Fig. 4a, the JP moved within more geometrically complicated optical patterns, such as circular, square and hexagonal rings. In all cases, the boundaries of the optical pattern on both sides of the JP defined and controlled its motion trajectory and hence, due to its tendency to remain within the optically defined boundary as was demonstrated above for the circular region (Fig.3) under sDEP mode, it remained within the edges. Similar control over the JP trajectory shape was obtained under ICEP mode (10kHz) of motion (see Fig. S5). This endows great operational flexibility in terms of JP guidance (e.g., to a desired destination or along a desired path), where instead of steering (e.g., magnetically) it using either open-[24,25] or closed-loop control[36], the desired path can be simply optically patterned. Such optical patterning can also replace either fabricated electrodes[37,38] or fabricated microchannels[36,39] that are inherently fixed and time- and resource-consuming.

This robust and directed, self-propelled movement of the JP within optically patterned regions was also demonstrated for its translocation between an array of isolated circular beams (Fig. 4b) by temporarily connecting these spatially separated regions by a straight optical line. This capability can be extended to control several JPs and to direct their assembly into an active structure, e.g., ring of JPs, as shown in Fig. 4c, while controlling the exact number of JPs participating in the self-assembly. Adjacent JPs interact in head-to-tail configuration (i.e., metallic side of one JP contacting the dielectric side of another JP) due to the dipolar attractive force acting between the opposite dipoles associated with the metallic and dielectric JP sides[27,40]. Thus, once a desired number of JPs is self-assembled into an active colloidal structure, the frequency can be tuned to control its angular velocity (i.e., counter-clockwise direction with the JP's dielectric side moving forward at 10kHz). In addition, the optical pattern can be modulated to a size comparable or even smaller in at least one of its dimensions of that of the colloidal structure, thereby, controlling its interaction with the optical pattern boundaries which may also lead to colloidal structure deformation (see Fig. S6). The same optoelectronic-directed self-assembly approach can be used to assemble a hybrid (passive and active particles) such as that depicted in Fig. S7, wherein a single passive particle is assembled into a hybrid active structure with a controlled number of JPs. The capability of the optoelectronically controlled JP to perform cargo (both synthetic made of 10μm polystyrene particles, and biological consisting of fixed 293T cancer cells of 12-15μm in size) loading, transport and delivery is demonstrated in Fig.S9 and movies S7, S8. Herein, the trajectory along which the JP acting as cargo carrier moved was optically defined. Previous studies have shown the biocompatibility of the optoelectronic platform and its usefulness for biological assays[21,22].

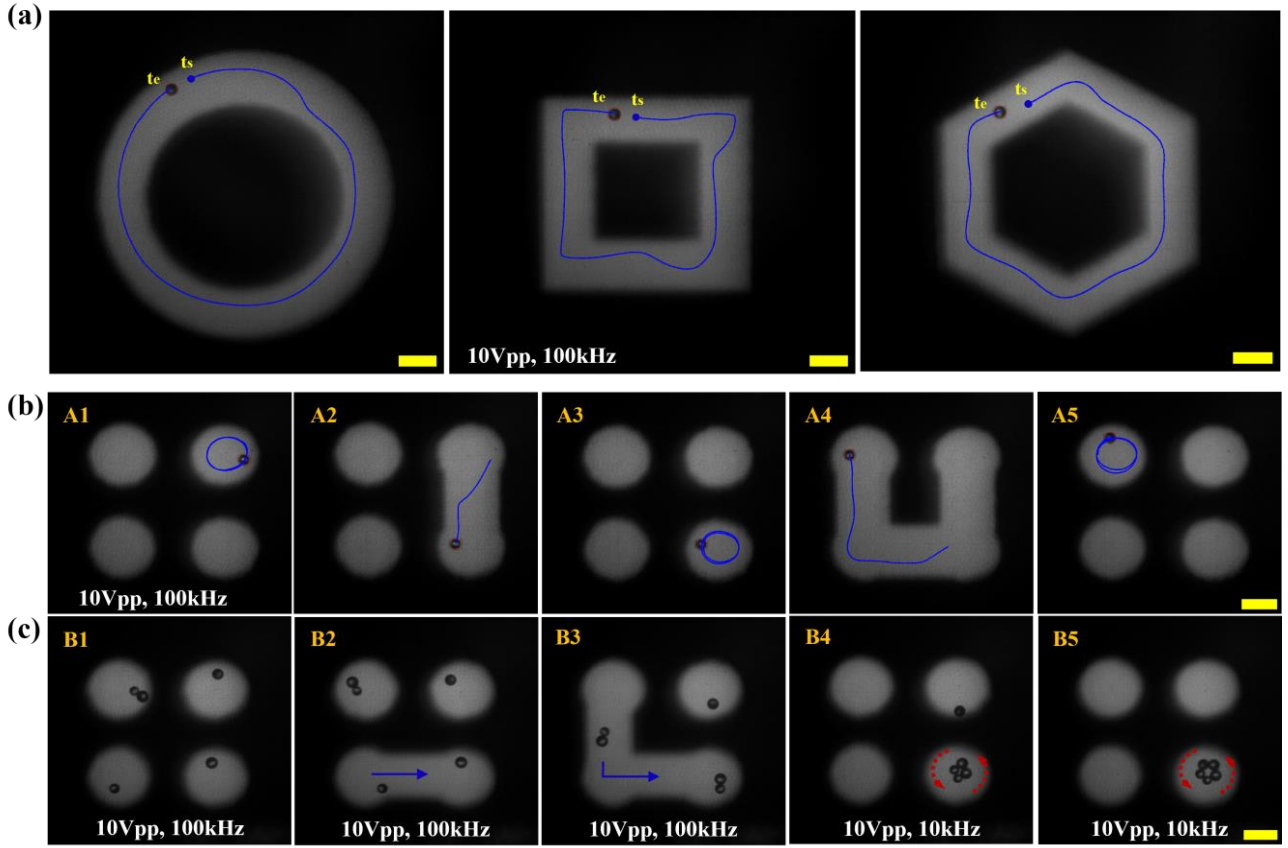

**Figure 4: Reconfiguration of a JP trajectory and directed self-assembly of JPs into an active colloidal structure.** (a) Motion of a JP within various geometrically (circular/square/hexagonal rings) patterned regions (applied field: 10Vpp, 100kHz). See also supplementary video S3. (b) A1-5: controlling the motion of a single JP between an array of optically patterned regions (applied field: 10Vpp, 100kHz). See also supplementary video S4; (c) B1-5: controlling multiple JPs individually to direct their self-assembly into an active colloidal structure (applied electric field: 10Vpp, 100kHz (B1-3) and 10Vpp, 10kHz (B4-5)). See also supplementary video S5. Scale bars: 100 μm.

*Dynamics of self-assembled structure of varying numbers of JPs*

Examination of the self-propelling behavior of the assembled structures consisting of a varying number of JPs within a circular optical region at the lower frequency (e.g., 10 kHz), found that the assembled structure approached and interacted with the edge of the optical region (Fig. 5). In most cases, the JPs first assembled in a chain configuration that moved linearly (in ICEP mode) towards the edge of the illuminated region, where the structure was perturbed, due to the locally intense electro-hydrodynamic (i.e., ACEO) flow, such that the JPs at the two ends of the chain approached each other to form a closed ring. The shape of the ring (e.g., triangular (3 JPs), square (4 JPs), pentagonal (5 JPs), hexagonal (6 JPs), heptagonal (7 JPs) etc.) (Fig. 5a, b) proved to be a function of the number of assembled JPs. The assembled closed ring of JPs then took on a rolling-like motion due to the continual interaction with the ACEO flow field that resulted in both linear motion along the edge of the optical region and angular motion around the assembly's centroid in the direction of motion of the dielectric side of the JPs (i.e., overall clockwise direction for the assemblies depicted in Fig. 5a, b). Both the linear and angular velocities depended on the size of the assembled colloidal structure and their magnitude monotonically increased with increasing numbers of assembled JPs (Fig.5b). In contrast to the single JP that was repelled from the optical edge at an applied electric field frequency of 10kHz (Fig.2a), the assembled (>3) JPs structures (Fig. 5) moved along the edge of the optical pattern, possibly due to the increased attractive pDEP force that scales with the effective increased assembly size[41].

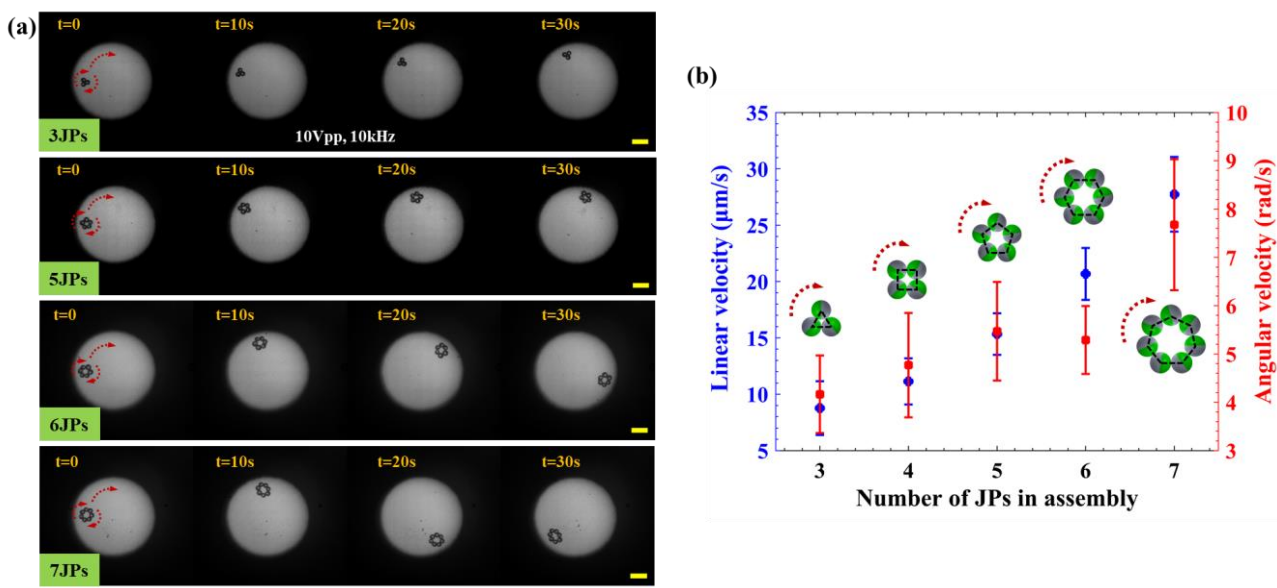

**Figure 5: Characterization of the propulsion of a self-assembled active structure of varying numbers of JPs.** (a) Shape of self-assembled colloidal structures and their motion characteristics with respect to the number of interacting JPs. Here the self-assembly rotational speed as well as propulsion velocity increased as the structure size increased. See also supplementary video S6. (b) Self-assembled structure propulsion linear and angular velocity vs. number of component JPs. Applied field: 10Vpp, 10kHz. Scale bars: 100 μm.

## Conclusions

This work demonstrated the ability to control the region within which JPs move by optically patterning electrodes using a DMD system for optical illumination of a photoconductive layer coating a conductive substrate. This extends previous studies where only a passive micromotor was optoelectronically manipulated. In contrast to the passive particle manipulation that necessitate phoretic forces via an optoelectronic tweezing effect, the active particles here were self-propelling and hence did not require an OET effect to phoretically translocate them. Instead, it required definition of the region of the optically patterned electrode within which the JPs move without crossing the edge of the optical region. Furthermore, a relatively low to highly geometrically constrained JP movement was achievable by modulating the size of the optical region. In addition, reduction of the size of the optical region comparable to that of the assembled structure of JPs resulted in its deformation. In addition, simultaneous manipulation of multiple JPs was demonstrated by patterning an array of electrodes which endowed additional control over directed self-assembly of JPs into stable structures (e.g., JPs ring) with precise control over the number of JPs participating in the assembly. Such assembled active structures also exhibited different propulsion behavior, with increasing angular and linear velocity within the optical region with increasing number of JPs. This strategy was further extended to demonstrate formation of hybrid assemblies consisting of multiple active JPs and a passive particle. We envisage that such an optoelectronic (OE) manipulation technique of active particles fosters enormous flexibility to manipulate several active particles simultaneously and direct their self-assembly into desired structures. In particular, the OE system is amenable for closed-loop operation based on real-time image analysis and thus enables transformation of these active micromotors into active microrobots and micromachines that can be programmed in real-time to perform various operations and assemblies in a highly parallelized and programmable manner, e.g., gene transfection of single targeted cells, cell fusion and cell separation.

## Methods and Materials

*Fabrication of JPs and solution preparation:* Polystyrene particles (diameter: 27 µm) (Sigma-Aldrich) in isopropanol (IPA) were pipetted onto a glass slide to form a monolayer upon solvent evaporation. The glass slide with particles was then coated with 15 nm-thick Cr, followed by 50 nm-thick Ni, and 15 nm-thick Au, as described by Wu et al.[42] To release the JPs, the substrate was sonicated in deionized water (DIW) with 2% (v/v) Tween 20 (Sigma Aldrich). Thereafter, the JPs were washed three times with the working medium to minimize adhesion to the substrate. The working solution was prepared by mixing 50 µM KCl (obtained from serial dilution of 1M KCl stock solution) and 0.1% (v/v) Tween-20 (conductivity ~ 8 µS/cm). For flow visualization and µPIV analysis, the solution was seeded with polystyrene tracer particles (diameter: 1 µm) (Fluoro-Max™) at a concentration of ~ 2% (v/v). The 293T cancer cells used for biological cargo manipulation demonstration (Fig.S9) were fixed with 4% formaldehyde (see Ref. [35] for details of the cell fixation procedure).

*Optoelectronic device fabrication and system operation:* The OE microfluidic chamber was fabricated by sandwiching a spacer layer (~120 µm-thick, Grace-Bio) between two parallel ~200-nm-thick indium tin oxide (ITO)-coated glass slides (Sigma-Aldrich) (see Fig. 1). The bottom ITO-coated glass slide was coated with ~800-nm-thick hydrogenated amorphous silicon (a-Si:H) using plasma-enhanced chemical vapour deposition (PECVD) (Oxford Plasma Instruments, Plasma Lab 100). In order to generate an a-Si:H layer, the following process recipe was used: $SiH_4$ (20 sccm), $H_2$ (60 sccm), substrate temperature of 200 °C, pressure of 1000 mTorr and RF power of 30W. The photoconductivity and dark conductivity of the fabricated a-Si:H layer was measured by monitoring the current under 5V that was applied between the ITO layer and the photoconductive layer illuminated by a ~5-mm-diameter beam, at an intensity of the order of ~$10^{-7}$ Sm$^{-1}$ and ~$10^{-10}$ Sm$^{-1}$, respectively. Copper tapes (Digi-Key, model: 3M9887-ND) were attached to the ITO-coated glass slides. The AC electrical field was applied between the two electrode leads connecting the copper tapes, using a function generator (Agilent, 33250A) and monitored by an oscilloscope (Tektronix, TPS-2024). The in-house-made OET system comprises a DMD-based pattern illuminator (Texas Instruments, DLP2000 EVM) integrated within a customized upright microscope (Olympus, model: U-ST) to enable illumination of light pattern onto the substrate (Fig. S1). The convex lens (Thorlabs, model: LB1757-A) in between the DMD illuminator and the microscope port focused the DMD-projected LED light (wavelength of ~625 nm, power ~0.23 W (or ~20 lumens) pattern onto the dichroic mirror which directs the incident light through a 10x objective lens onto the OET substrate placed on a XY manual stage (Thorlabs, model: PT1). The DMD illuminated pre-determined stationary image patterns. Imaging was performed using a CMOS camera (Thorlabs, model: DCC1645C) connected at the microscope's top port, through a short-pass filter (Thorlabs, model: FES0550, cut-off wavelength ~550 nm) to avoid camera pixel saturation by the high intensity of reflected red beam from the substrate. The optical power density of the circular light pattern projected through a 10x objective lens measured ~0.051 W/cm$^2$ at the substrate plane, as determined using an Ophir Photonics VEGA power meter. The quadrupolar electrode array (Fig.S4) used for characterization of the JP's DEP response was fabricated using standard photolithography techniques following metallic deposition of 15nm Cr and 200nm Au[41].

*Image analysis:* All images were analyzed using Python (3.9.7) and ImageJ software. Microparticle image velocimetry (μPIV) analysis was performed using PIVLab (Matlab 2018), where the flow field of 100 images was averaged to evaluate the velocity field data. The velocity magnitude was then evaluated based on the calibration scale of the real system and time step between each image frame.

*Numerical simulations:* The numerical simulation of the electrostatic and electro-hydrodynamic problem was performed in COMSOL™ 5.3. A simple 2D geometry, consisting of a rectangular chamber of 120 μm in height and 240 μm in width, with the top surface acting as a conductor while only a region of 180 μm of the bottom surface was conducting, was used to model the small beam (180 μm in diameter) region depicted in Fig. 3. In the analysis, which followed that of Green et al.[42,43], it is assumed that electrolysis does not occur at the electrode surfaces and the electrodes may be considered as ideally polarizable. Additionally, in accordance with the weak field assumption wherein the applied voltages are small relative to the thermal potential, any surface conduction or convection within the electric double layer (EDL) can be neglected, so that the double layer may be modelled as a linear capacitor. Since the EDLs are thin relative to the height of the chamber $\left(\frac{\lambda}{H} \ll 1\right)$ we can solve the Laplace equation for the electric potential, $\phi$, in conjunction with the following boundary condition at the electrode surface

$$\sigma \frac{\partial \phi}{\partial n} = i\omega C_{DL}(\phi - V_j), \quad (1)$$

which describes the oscillatory Ohmic charging of the induced EDL in response to an AC potential applied at the electrodes of magnitude $V_j$ ($j$ is the electrode's index), and angular velocity $\omega$. Herein, $n$ is the coordinate in the direction of the normal to the electrode surface, and $C_{DL}$ represents the capacitance per unit area of the EDL and can be estimated from the Debye-Huckel theory as $C_{DL} \sim \varepsilon/\lambda$ (neglecting the Stern layer), where $\varepsilon$ is the (real) permittivity of the medium. A voltage of $V_0$ was applied on the upper wall of the chamber, while the optically patterned region of 180 µm on the lower wall was grounded. An insulation boundary condition was applied on the other substrate surfaces (see Fig. S3).

Since the electrostatic and hydrodynamic equations are decoupled ($Pe = 0$), based on the thin EDL approximation, the velocity field can be obtained using the unforced Stokes equation, accounting for the electric forcing by simply prescribing an effective slip velocity at the electrodes. The time-averaged slip velocity boundary condition on the electrodes is[43,44]

$$\langle u \rangle = -\frac{\varepsilon}{4\eta} \frac{\partial}{\partial x}\left[(\phi - V_j)(\phi - V_j)^*\right], \quad (2)$$

wherein the symbol * indicates the complex conjugate. At the other physical boundaries, no slip boundary conditions were used since the linear electroosmotic flow contribution has a zero time-average effect under AC forcing.

## Acknowledgments

G.Y. acknowledges support from the Israel Science Foundation (ISF) (1934/20). Fabrication of the chip was made possible through the financial and technical support of the Russell Berrie Nanotechnology Institute and the Micro-Nano Fabrication Unit. We thank Dr. Michael Shustov for his help in setting up the experimental DMD system and Dr. Yue Wu for her advice and guidance on using electric fields for propelling JPs. We also thank Prof. Eric Chiou from the University of California at Los Angeles, USA for his advice regarding photoconductive substrates. In addition, we thank the Center for Nanoscience and Nanotechnology, Hebrew University of Jerusalem for assisting us in fabricating the photoconductive substrates.

## Conflict of Interest

The authors declare no conflict of interest.